\documentclass{aastex}
\usepackage{graphicx}
\usepackage{float}
\usepackage{longtable}

\title{A 3-dimensional model of tangential YORP\altaffilmark{1}}
\altaffiltext{1}{Dedicated to peace in Ukraine}
\author{O. Golubov\altaffilmark{2}}
\altaffiltext{2}{Also affiliated with Institute of Astronomy of V. N. Karazin Kharkiv National University}
\author{D. J. Scheeres}
\affil{Department of Aerospace Engineering Sciences, University of Colorado at Boulder}
\affil{429 UCB, Boulder, CO, 80309, USA}
\email{\textbf{golubov@astron.kharkov.ua}}
\and
\author{Yu. N. Krugly}
\affil{Institute of Astronomy of V. N. Karazin Kharkiv National University}
\affil{35 Sumska Str., Kharkiv, 61022, Ukraine}

\begin{abstract}
Tangential YORP, or TYORP, has recently been demonstrated to be an important factor in the evolution of an asteroid's rotation state.
It is complementary to normal YORP, or NYORP, which used to be considered previously.
While NYORP is produced by non-symmetry in the large-scale geometry of an asteroid,
TYORP is due to heat conductivity in stones on the surface of the asteroid.
Yet to date TYORP has been studied only in a simplified 1-dimensional model, substituting stones by high long walls.

This article for the first time considers TYORP in a realistic 3-dimensional model, also including shadowing and self-illumination effects via ray tracing.
TYORP is simulated for spherical stones lying on regolith.
The model includes only 5 free parameters,
and the dependence of the TYORP on each of them is studied.
The TYORP torque appears to be smaller than previous estimates from 1-dimensional model, but still comparable to the NYORP torques.

These results can be used to estimate TYORP of different asteroids,
and also as a basis for more sophisticated models of TYORP.

\end{abstract}


\begin{document}

\section{Introduction}
The YORP effect is a torque acting on an asteroid, 
created by the recoil force of the light reflected or reemitted by the surface (\cite{rubincam00}, \cite{bottke06}).
In simulations of YORP the heat conductivity in the asteroid used to be considered as 1-dimensional, 
and the local curvature of the asteroid's surface was neglected \citep{rozitis13}.
Under such assumptions, the YORP torque was only due to non-symmetries of the asteroid's shape,
while a symmetric asteroid could possess no YORP.
Heat conductivity through the asteroid's body was accounted only for small asteroids measuring some metres in diameter \citep{breiter10}.
Still, surfaces of asteroids can be covered with stones, making the heat conductivity problem at the surface 3-dimensional and allowing heat fluxes through the stones.

Recently \cite{golubov12} demonstrated that accounting for these effects can substantially change the picture of YORP.
Heat fluxes through stones on the asteroid's surface can cause their western sides to be slightly warmer than their eastern sides,
thus causing them to experience a net drag parallel to the global surface of the asteroid, and to create a torque increasing the rotation rate of the asteroid.
This torque was called the tangential YORP (or TYORP), in contrast to the normal YORP (or NYORP),
which had been considered previously and is produced by forces normal to the global surface.
Even a perfectly symmetric asteroid was demonstrated to experience TYORP, while its NYORP is nill.
For reallistic moderately asymmetric asteroids the strength of TYORP torque was estimated as comparable to the one of NYORP.

Still, these estimates were very rough, as \cite{golubov12} did all their simulations in a simple 1-dimensional model,
substituting stones with high thin walls standing on the asteroid's surface in the meridional direction.
Thus their consideration gave just an order-of-magnitude estimate of the strength of TYORP,
and the question of a more precise description of the effect remained.

In this article we construct a more realistic model of TYORP, and study an asteroid's surface covered with spherical stones.
In Section 2 we describe our model and review the methods used for its simulations.
Results of the simulations are presented in Section 3.
In Section 4 we discuss the results and their implications.
In Appendix A discuss our numeric algorithm in more detail,
and in Appendix B we derive formulae for the integrated TYORP torque of an ellipsoidal asteroid.

\section{Model}

\begin{figure}
  \centering    
  \includegraphics[width=150mm]{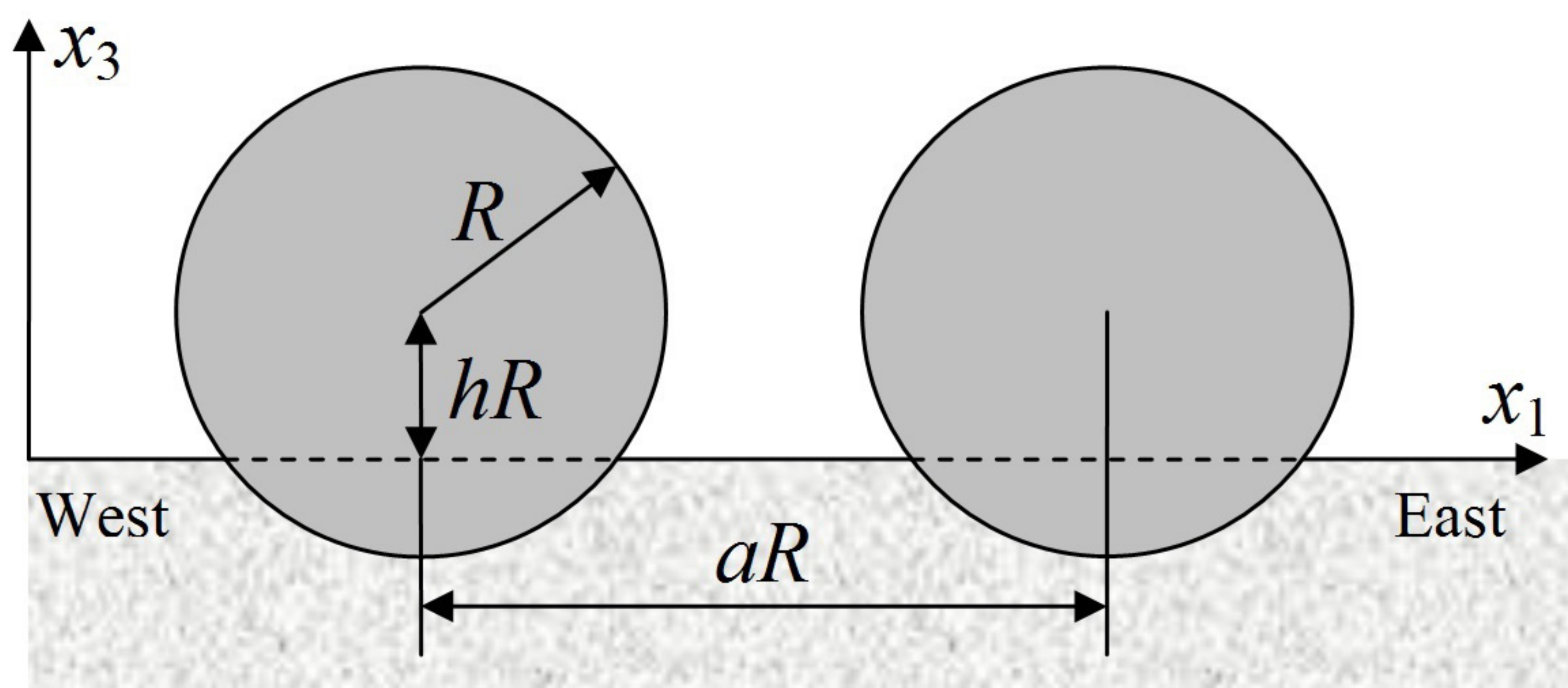}
  \caption{Model studied in the article. Spherical stones of radius $R$ lie on regolith.
  Centers of stones are a height $hR$ above the level of regolith. The distance between the stones is $aR$.
  Heat conductivity of the stones is characterized by the heat parameter $\theta$, the regolith is a perfect heat insulator.}
  \label{spheres}
\end{figure}

We consider a flat patch of the asteroid's surface covered with spherical stones, as shown in Figure \ref{spheres}.
Each stone has radius $R$, and its center is situated at a height $hR$ above the surface ($-1<h<1$).
Stones are arranged in a periodic square grid of size $aR$ ($a \ge 2$), with the sides of the squares going in the directions south-north and east-west.
The size of the patch under consideration is assumed to be much smaller than the size of the asteroid,
so that we disregard curvature of the surface.
We describe the entire patch with the same latitude $\psi$, determined as the angle between the normal of the patch and the asteroid's equatorial plane.
The entire semispace below the surface ($z_3 < 0$) is filled with regolith, so that there is regolith everywhere between and under the stones.
The heat conductivity of the regolith is assumed to be much smaller than the heat conductivity of stones,
so that no heat conductivity between a stone and the surrounding regolith occurs,
and the reemission of the absorbed heat by the regolith is instantaneous.

The heat conductivity of the stones is $\kappa$, the heat conductivity of the regolith is 0.
The albedo of both the stones and the regolith is $A$.
The heat capacity of a stone is $C$ and its density is $\rho$.
Then the temperature distribution in a stone obeys the heat conductivity equation
\begin{equation}
 C\rho\frac{\partial T}{\partial t}=\kappa\sum_{i=1}^3\frac{\partial^2 T}{\partial x_i^2}.
 \label{conductivity}
\end{equation}
The boundary condition for this equation above the ground is
\begin{equation}
 \kappa\frac{\partial T}{\partial x_i}=\left\{ 
  \begin{array}{c c}
    n_i((1-A)\alpha\Phi-\epsilon\sigma T^4), & x_3>0,\\
    0, & x_3 \le 0.
  \end{array} \right.\
  \label{boundary}
\end{equation}
Here $\sigma$ is Stefan--Boltzmann's constant, $\epsilon$ is emissivity of stone, $\boldmath{n}$ is the normal vector of the surface,
and $\alpha\Phi$ is the incoming light power per unit surface of the stone, 
with $\Phi$ being the solar constant and some variable coefficient $\alpha \la 1$.

The characteristic scale of the temperature is the equilibrium temperature of the subsolar point,
\begin{equation}
 T_0=\sqrt[4]{\frac{(1-A)\Phi}{\epsilon\sigma}}.
\end{equation}
Distance has two important scales, namely the wavelength of the heat conductivity wave $L_\mathrm{wave}$, 
and the heat conductivity length $L_\mathrm{cond}$, expressed by formulae
\begin{equation}
 L_\mathrm{wave}=\sqrt{\frac{\kappa}{C\rho\omega}},
 \label{L_cond}
\end{equation}
\begin{equation}
 L_\mathrm{cond}=\frac{\kappa}{\left((1-A)\Phi\right)^{3/4}\left(\epsilon \sigma \right)^{1/4}}.
\end{equation}
The physical meaning of $L_\mathrm{cond}$ is that it is the distance 
at which temperature difference equal to $T_0$ causes heat flux equal to $\Phi$.
Usually at distance scales much bigger than $L_\mathrm{cond}$, heat conductivity can be neglected.
The ratio of $L_\mathrm{cond}$ and $L_\mathrm{wave}$ is called the thermal parameter
\begin{equation}
 \theta=\frac{\left(C\rho\kappa\omega\right)^{1/2}}{\left((1-A)\Phi\right)^{3/4}\left(\epsilon \sigma \right)^{1/4}}.
 \label{theta}
\end{equation}
The thermal parameter characterizes the relative importance of heat conductivity with respect to heat absorption and emission.

We non-dimensionalize all the variables.
We introduce dimensionless variables $\xi=x/L_\mathrm{cond}$ and $\tau=T/T_0$.
Instead of time $t$ we use the rotation phase $\phi=\omega t$, with $\omega$ being the angular velocity of the asteroid.

In these terms Equations \ref{conductivity} and \ref{boundary} transform into
\begin{equation}
 \frac{\partial \tau}{\partial \phi}=\frac{1}{\theta^2}\sum_{i=1}^3\frac{\partial^2 \tau}{\partial \xi_i^2}.
 \label{conductivity_nonD}
\end{equation}
The boundary condition for this equation above the ground is
\begin{equation}
 \frac{\partial \tau}{\partial \xi_i}=\left\{ 
  \begin{array}{c c}
    n_i(\alpha-\tau^4), & \xi_3>0\\
    0, & \xi_3 \le 0
  \end{array} \right.\
  \label{boundary_nonD}
\end{equation}

To simplify the analysis we keep the number of free parameters to the minimum.
In our dimensionless simulations we assume albedo $A=0$, emissivity $\epsilon=1$, and Lambert's law for scattered and emitted light.
Allowing for different $A$, different $\epsilon$, different scattering and emission laws,
would make the problem too difficult to tackle.
If no instances of self-illuminations occur,
i.e. a ray emitted by the asteroid never falls back onto the asteroid,
the non-dimensionalized results can be easily rescaled to different $A$ and $\epsilon$ with the aid of Equations \ref{L_cond} and \ref{theta}.
And there is a good reason to believe that this rescaling gives relatively accurate results even in the presence of some self-illumination,
as on the one hand usually only a minor portion of emitted or reflected light falls back onto the surface, on the other hand $A$ is usually only slightly deviates from 0 and $\epsilon$ only slifhtly deviates from 1,
so accounting for both these effects simultaneously should only give a second order correction.
Without investigating this question in more detail,
we non-dimensionalize our model with Equations \ref{L_cond} and \ref{theta}, and then in the non-dimensional model assume $A=0$ and $\epsilon=1$.
This treatment is precisely correct if in the initial model really $A=0$ and $\epsilon=1$,
and if not then this treatment presumably gives a good approximation.

We simulate the heat conductivity in stones and the ray tracing numerically
using Monte Carlo technique.
The stone is modelled with the aid of a finite difference method on a cubic mesh.
We assume all stones to be the same, thus posing periodic boundary conditions
and assuming that a ray leaving through the left boundary reappears on the right boundary.
We cast rays from the Sun onto the asteroid and trace each of them.
If a ray is absorbed by the regolith on the surface, it is instantly reemitted with the same energy
and with a random direction determined in accordance with Lambert's law.
If a ray hits a stone, it is absorbed and its energy is deposited to the closest node of the mesh within the stone.
Then the stone emits rays according to the Stefan--Boltzmann's law.
The directions of the rays are chosen randomly according to Lambert's law,
their initiation points are randomly chosen on the stone's open surface,
and their energies are determined by temperatures in the closest nodes of the mesh.
The energy of each ray is subtracted from the neighbouring node.
Then the ray is traced, and is either absorbed by another stone (and then its energy is returned back) or goes into space. 
It can be also scattered by the regolith.
We repeat the procedure many times, as the Sun follows its diurnal path.
The simulation continues for several asteroidal days to make the system forget its initial conditions.
Then the momentum in $x$ direction given to asteroid by the emitted rays is calculated and averaged over several days.
Our numeric algorithm is explained in more detail in Appendix A.

The momentum is expressed in dimensionless units as the force acting on a stone
divided over the area occupied by the stone and over the solar light momentum flux,
\begin{equation}
 p_x=\frac{F_x c}{\pi R^2 \Phi}.
 \label{px_normalization}
\end{equation}
If we want go from $p_x$ to the force per unit area $P$,
we must multiply $p_x$ by the solar light momentum flux $\Phi/c$
and by the fraction of the surface area occupied by the stones.

The trade-off between parameters governing the simulation (the size of the spatial mesh, the timestep, 
the number of rays cast from the Sun and of rays emitted by the stone in each step, the number of asteroidal days simulated)
is adjusted so that we reach the highest accuracy for a given computation time.
Then $p_x$ is studied as a function of physical parameters.

\begin{figure*}
  \centering    
  \includegraphics{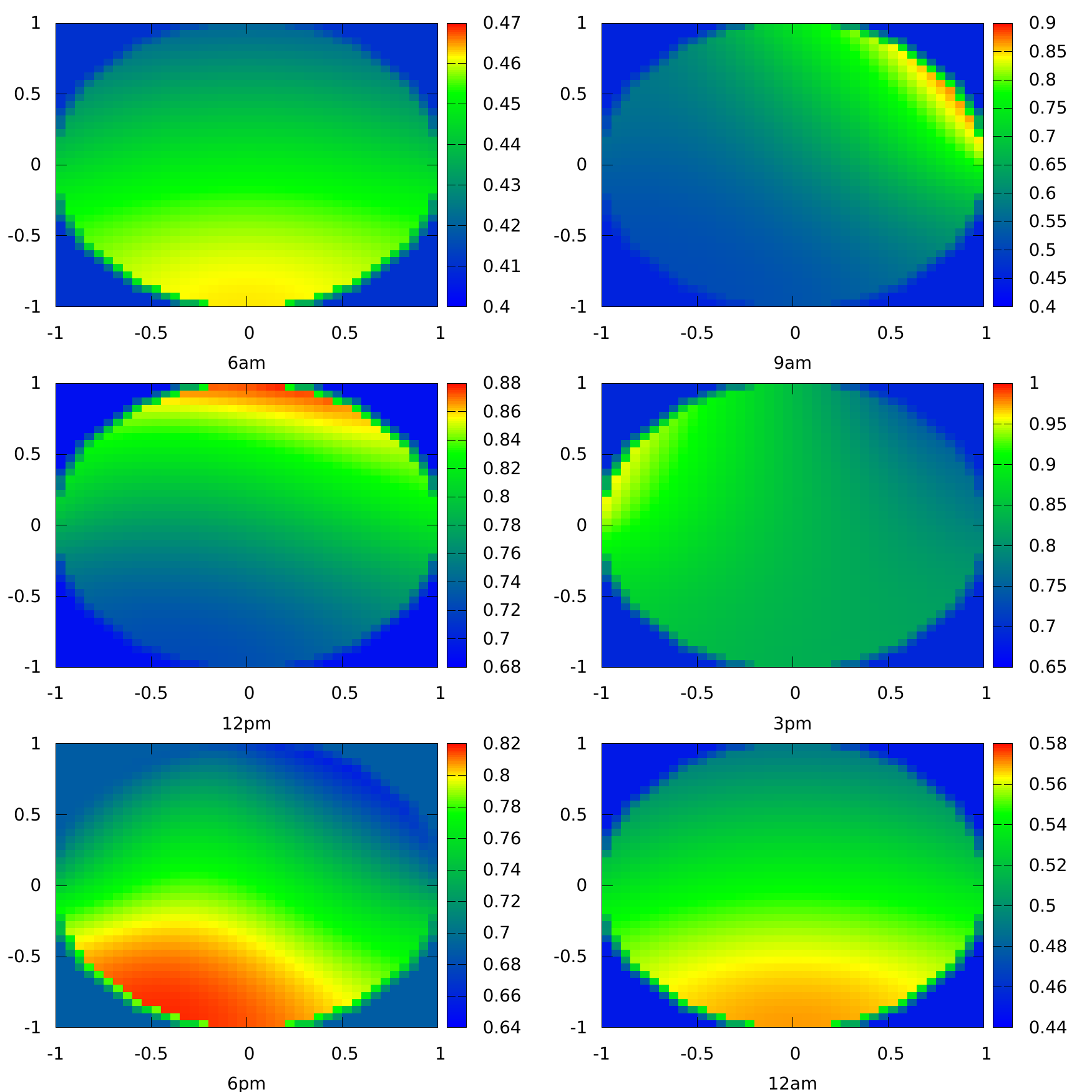}
  \caption{Temperature distribution in the stone at different instants of time.
  Each of the six panels shows temperature distribution in the East-West cross-section passing through the centre of the stone,
  and has the same orientation as Figure \ref{spheres}.
  Time is marked under each panel and measured in ``asteroid hours'', so that 12 denotes midday, and 0 and 24 denote midnight.
  Parameters used for the simulation are $r=1$, $a=3$, $h=0$, $\psi=0$, $\theta=1$,
  temperature is expressed in dimensionless units $\tau=T/T_0$.
  Note different temperature scales for different panels.}
  \label{tau}
\end{figure*}

\begin{figure}
  \centering    
  \includegraphics{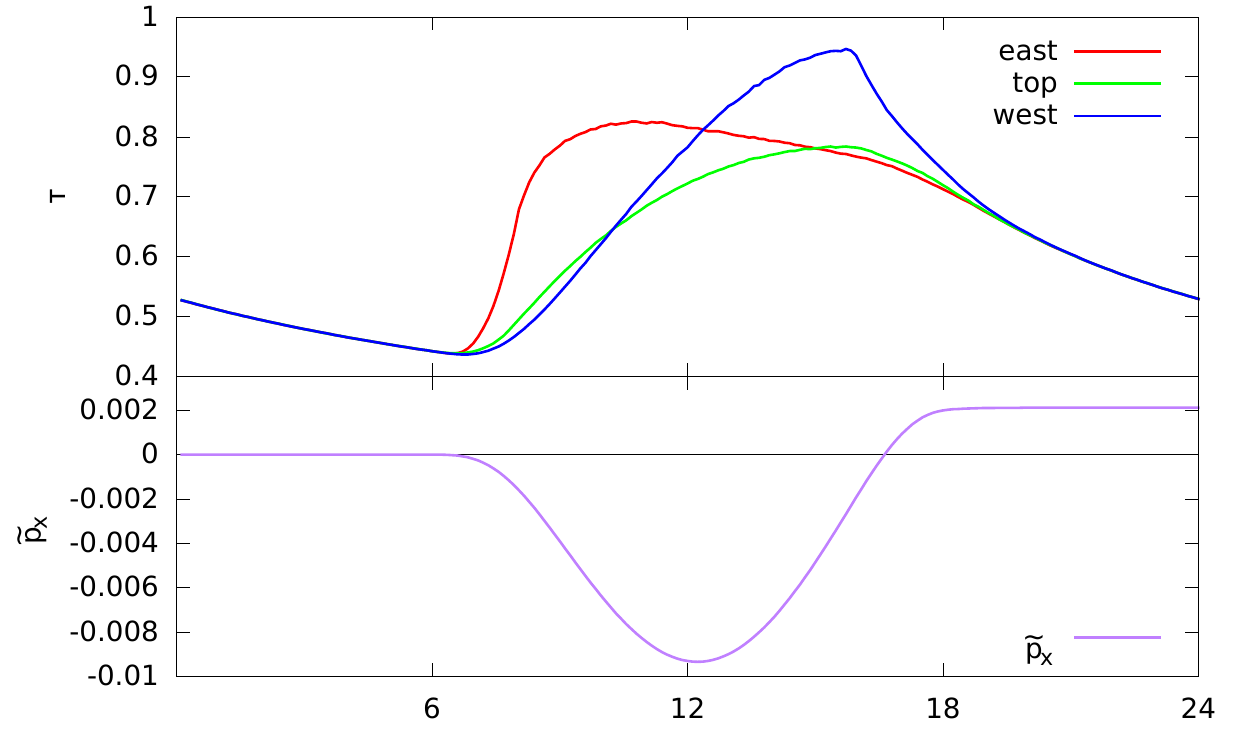}
  \caption{Temperature in the stone and TYORP drag force as functions of time.
  Time is expressed in ``asteroid hours''.
  The upper panel shows temperatures in the eastmost, top, and westmost points of the stone.
  Temperatures in the eastmost and westmost points differ significantly.
  The lower panel shows different scale we plot the time-integrated $p_x$.
  When the eastern part of the stone is warmer than the western part $p_x$ decreases,
  when it is cooler $p_x$ rises, and in the end $p_x$ reaches a positive value.}
  \label{tau1D}
\end{figure}

Origination of the effect can be explained with the aid of Figures \ref{tau} and \ref{tau1D}.
Figure \ref{tau} shows the temperature distribution inside the stone at six different moments.
At sunrise (6am) the temperature of the stone is the lowest, as the asteroid was cooled all night long.
In the morning the stone is heated from the East, at noon from the top, and in the afternoon from the West,
that is illustrated in the next 3 panels.
At 6pm sun sets, heating stops and the stone slowly cools down.
In the lower left panel the surface of the stone is already relatively cool 
as for some time before 6pm the stone has been largely shadowed by the next stone to the West.
The western side of the stone is still significantly warmer than its eastern side,
but this difference vanishes before midnight.

In Figure \ref{tau1D} we plot the temperatures of the three most characteristic points of the stone: 
the eastmost, the top, and the westmost points.
We see that the temperature in the East is the first to start rising.
It pulls up the temperatures of the other points due to heat conductivity.
When the afternoon sun starts heating the western part of the stone, its temperature rises even more,
and reaches levels never attained by the eastern part of the stone.
Even though the mean temperature for the two parts is nearly the same,
the temperature in the East is more uniform, while the temperature in the West has a sharper maximum.
As a result the mean fourth power of the temperature is bigger in the West, so is the recoil force due to the Stefan-Boltzmann's law.

In the lower part of the figure we plot the TYORP force integrated over time (starting at midnight),
\begin{equation}
 \tilde{p}_x(t)=\frac{1}{t_{rot}}\int_0^t\frac{F_x(t_1) c}{\pi R^2 \Phi}\, \mathrm{d}t_1.
\end{equation}
We see that in the morning the integral gets negative, as the eastern side of the stone is warmer, 
emits more infrared light, and slows down rotation of the asteroid.
In the afternoon the western side of the asteroid gets warmer, causes the integrated TYORP force to increase, and in the end prevails.
The integral of the TYORP force over the whole rotation period is positive,
which means a positive $p_x=\tilde{p}_x(t_\mathrm{rot})$.

\section{Results}
In Figure \ref{p_x} we plot the dimensionless TYORP force $p_x$ as a function of different parameters.
There are 5 relevant parameters:
dimensionless radius of the spherical stones $r$,
relative distance between the stones $a$,
relative height of the center of a stone above the ground $h$,
latitude $\psi$, 
and thermal parameter $\theta$.
Five panels of the plot show dependencies on one variable each,
and only the first panel shows a 2-dimensional dependence on $r$ and $\theta$ colour-coded.

From the first panel we see that the TYORP acceleration is significant only inside an ellipsis stretching from the upper left to the lower right.
The maximum is attained around $r=0.3$ and $\theta=2$, where we have $p_x \approx 0.003$.
The function decreases from this point to the lower left and the upper right very steeply,
while the decrease in the upper left and the lower right is much slower.

The middle and the lower panel in the left show cross-sections of the upper left panel 
in the horizontal and vertical lines respectively.
We see maxima in each line, situated in the area where the line crosses the red ellipsis in the upper left panel.
The maximum is lower if the intersection point is farther from the centre of the ellipsis.

The general appearance of the upper and the lower left panels is similar to the lower two panels in Figure 2 in \cite{golubov12}.
The physical interpretation is also similar;
the TYORP effect originates as a result of the heat conductivity lag on the scale of the stone,
therefore for any substantial effect, the heat conductivity length, the thermal wavelength, and the size of the stone must be comparable.

In the upper right panel Figure \ref{p_x} we plot the dimensionless TYORP force $p_x$ 
as a function of the latitude $\psi$.
We see that $p_x$ is biggest at the equator of the asteroid ($\psi=0$), and goes to 0 at the poles ($\psi=90^\circ$).
Dashed lines showing a cosine function are overplotted for comparison.

The middle right panel shows the dependence of $p_x$ on the distance between the stones $a$.
The left limit of the plot $a=2$ corresponds to stones touching each other.

The last panel shows $p_x$ as a function of the height of the center of a stone above the ground $h$.
When $h$ is close to $-1$, the stones are almost entirely below the ground, and the effect vanishes.
Then when $h$ increases to 0, $p_x$ also increases.
Finally, when $h$ goes on increasing to 1, $p_x$ does not demonstrate any more significant increase,
due to shadowing of the lower parts of the spheres.

\section{Discussion}

To estimate the relative importance of TYORP and NYORP we follow \cite{golubov12} and use the normalized YORP torque, determined as
\begin{equation}
 \tau_z=\frac{T_z c}{\Phi r_\mathrm{eq}^3},
 \label{tau_z}
\end{equation}
with $T_z$ being the YORP torque, $r_\mathrm\mathrm{eq}$ being the equivalent radius of the asteroid (the radius of the sphere of the same volume),
$c$ being the speed of light, and $\Phi$ being the solar radiation flux at the position of the asteroid.
Estimate of dimensionless NYORP give $\tau_z=0.008$ for 1620 Geographos \citep{durech08geo} and $\tau_z=0.002$ for 54509 YORP \citep{lowry07}.

To estimate $\tau_z$ we need to integrate the TYORP force over the surface of the asteroid.
As the latitude dependence of $p_x$ (upper right panel of Fugure \ref{p_x}) is relatively complicated,
we try two limiting cases, a sinusoidal dependence and a constant (see Appendix B).
Both estimates give nearly the same result,
\begin{equation}
 \tau \approx 9 p_0 f,
\end{equation}
where $p_0$ is $p_x$ at equator and $f$ is the fraction of the surface occupied by stones.
It implies that in our model the dimensionless TYORP can reach up to about 0.01 (for $p_0$ about 0.003 and $f$ close to 0.5).
This is an order of magnitude less than the value obtained by \cite{golubov12}, which makes sense 
as now the presence of the upper boundary of stones as an emitter and absorber of heat must decrease the temperature contrast in the stone,
and the presence of underground part of the stone acting as a heat reservoir must also level temperature gradients.
Still, this value is comparable to the strength of NYORP.

If we have spherical stones of different sizes lying on the surface of an asteroid,
then the TYORP force acting on a patch of the surface should be integrated over all sizes of stones.
The problem is especially complicated because of effects of shadowing and self-illumination,
so that stones can not be considered independently, but only in toto.
It is evident from the middle right panel of Figure \ref{p_x}.
If stones were not influencing each other, the total force would be proportional to their number,
and $p_x$ would be constant.
It eventually happens at $a\rightarrow \infty$, where $p_x$ reaches saturation.
For such low surface densities of stones they stop influencing each other,
and the overall TYORP force can be obtained as the integral over all sizes of stones 
of particular TYORP force for each size calculated in the absence of any other stones.
But for $a \approx 2$, where stones lie close to each other, significant deviation from the saturated limit are observed.
Moreover, $p_x$ in this case is sensitive to the arrangement of stones,
so that our results obtained for a square grid are only an estimate of what happens if stones are positioned more randomly.
In any case, significant contribution to TYORP is given only by stones of some particular sizes,
belonging to the maximum in the upper left panel of Figure \ref{p_x}.

The observed and predicted TYORP acceleration of Itokawa is compared in Table \ref{tab-itokawa}.
Despite a large variation between different predictions, they are systematically smaller than the observations by few $10^{-3}$.
\cite{golubov12} assumed that this descrepancy could be due to TYORP, but could support this claim only with rough estimates.
Now we are capable of a much more quantitative analysis.
We estimate TYORP for Itokawa, assuming rotation period 12.1 hr, semimajor axis 1.324 AU, $\kappa = 2.65$ W m$^{-1}$K$^{-1}$,
$C = 680$ J kg$^{-1}$K$^{-1}$, $\rho = 3500$ kg m$^{-3}$, $A=0.23$, $\epsilon = 0.7$.
For such parameters we get $\theta = 18$, $L_\mathrm{cond}=1.5$ m, $L_\mathrm{wave}=9$ cm.
The dimensiomless TYORP drag $p_x$ reaches the maximal value of 0.00025 for stones with radius $R\approx 4$ cm,
This maximum is very broad, so that $p_x > 0.00015$ for $R=1\div 15$ cm.
The corresponding dimensionless TYORP torque is $\tau_z\approx 0.002 f$,
where $f$ is the fraction of the surface occupied by stones with radii of 1 to 15 cm.
If these stones are really abundant on the surface ($f\approx 0.5$), TYORP can suffice to account for the descrepancy between the theory and the observations.
But even if the stones of these sizes are relatively rare ($f\approx 0.1$), TYORP still must provide a major contribution to the observed YORP acceleration.

If $\theta$ is not 18 as for Itokawa, but an order of magnitude less, TYORP can be an order of magnitude bigger.
This could happen for slow rotators, so that the slower the rotatation becomes the bigger is TYORP to speed up the rotation.
It could be the reason why slow rotators are rarely observed.

If we have any asteroid of known shape,
for each patch of whose surface the matelial properties of stones 
and their size distribution is known,
the results of this article can be used to reliably estimate the TYORP torque for this asteroid.
We must only add up TYORP forces produced by stones of different sizes of each patch of the surface,
and then integrate these forces over the whole surface of the asteroid to get the torque.
Limitatios of this method include: non-sphericity of stones;
non-convexity of the overall shape of the asteroid, which will alter illuminations conditions of some patches;
mutual shadowing and self-illumination of stones,
which can not be precisely accounted for if the stones are not arranged in a regular pattern.
Still, methods similar to the ones used in this article can be applied
for any shapes and mutual distributions of stones, and illmination conditions can also be adjusted to account for shadowing of one part of the asteroid by others.
This problem will depend on a very big number of free parameters.
It is hard to tackle this problem in a general case,
but it can be solved individually for each asteroid of interest and with enough data.
And even if such detailed data are available and such a sophisticated simulation is performed, the results of this article can be useful as a simple and robust estimate of the TYORP torque.

\section{Acknowledgements}
OG is very greateful to Dr. Anton Tkachuk for helpful discussion of numerical methods and revising Appendix A, 
to Dr. Glib Ivashkevych for speeding up the program,
and to Prof. Cornelis P. Dullemond for discussing the algorithm implemented in the program.
OG and DJS acknowledge support from NASA Grant NNX11AP24G.

\appendix
\section{Numerical methods}
In this appendix we describe in more detail the numerical algorithms used to solve Equations \ref{conductivity_nonD} and \ref{boundary_nonD}.

We separate each stone into small cubes of the size $\mathrm{d}r=r/N_r$, 
and discretize dimensionless time $\phi$ into intervals $\mathrm{d}\phi=1/(sN_r^2)$.
Here $N_r$ and $s$ are some constants, which have to be big enough to provide a good accuracy of the solution.
In each timestep $\mathrm{d}\phi$ we 
1) trace the incoming rays and add the energy brought by them to the surface of the stone, 
2) do one step of the heat conductivity equation,
3) subtract the emitted energy from the surface and trace the outcoming rays.
All simulations are done within an area of $ar\times ar$ assuming that all neighbouring stones have the same properties.

1) We run $N_\mathrm{vis}$ rays. All rays come from the sun and thus have the same direction, 
but initial coordinates are random so that the rays uniformly cover the area $ar\times ar$.
Each ray brings in the energy $\mathrm{d}E=-a^2r^2\sin \phi\cos \Psi\,\mathrm{d}\phi/N_{rays}$.
If the ray is absorbed by the stone, the closest node to the absorption point is found, 
and its temperature is increased by $\mathrm{d}E/(\theta^2\mathrm{d}r^{3})$.
If it is absorbed by regolith, it is instantly re-emitted from the same point, with the direction determined by Lambert's law.
If the ray leaves through the side of the simulated volume, it reappears on the other side with the same direction. 
These periodic boundary conditions imply periodic arrangement of similar stones in a square grid on the surface.
If the ray leaves the simulated volume through the top, it stops being calculated.

2) The heat conductivity equation is solved using the first order explicit finite-difference scheme in all three directions consequently.
The scheme is chosen for its simplicity of realization and fast performance.
As computation errors and shot noise from the ray tracing deteriorate the accuracy,
it makes no sense to implement a more sophisticated scheme.

3) We run $N_\mathrm{IR}$ infrared rays emitted by the stone.
The emission point of each ray is chosen at random on the open surface of the stone,
the direction of the ray is chosen at random in accordance with Lambert's law,
and the energy is proportional to the fourth power of the temperature of the nearest node.
The temperature of this nearest node is decreased in accordance with the energy taken away by the ray.
The ray is traced in the manner similar to the step (1), with the possibility of being returned back to the stone.
Simultaneously with each ray we trace the ray symmetric to it with respect to the vertical axis crossing the centre of the stone.
(This allows us to reduce the shot noise produced by the limited number of rays.)

As the initial condition we set a uniform temperature distribution inside the asteroid,
with the temperature being equal to the mean temperature at the latitude of the stone.
Then we study the evolution of the temperature for $2t_\mathrm{eq}$ rotation periods.
The first $t_\mathrm{eq}$ periods are not used to compute the TYORP force. 
They are introduced only to give enough time to the stone to forget the initial conditions.
The TYORP force is computed as the average over the last $t_\mathrm{eq}$ rotation periods.

We implement this algorithm in a program written in C++.
The program uses only standard libraries $cmath$ and $cstdlib$.
All procedures related to the ray tracing and integration of the heat conductivity equations are written from scratch.

The program requires 5 physical parameters of the stones ($r$, $\theta$, $h$, $a$, $\psi$),
and also 5 simulation parameters of the algorithm ($N_r$, $N_\mathrm{vis}$, $N_\mathrm{IR}$, $s$, $t_\mathrm{eq}$),
which must be adjusted depending on the physical parameters and the available computation time to provide the best possible accuracy.

The computation time is roughly proportional to
\begin{equation}
 t_\mathrm{comp} = C_1 t_\mathrm{eq}sN_r^2(N_r^3+0.46N_\mathrm{vis}+1.9N_\mathrm{IR}),
 \label{t_comp}
\end{equation}
where $C_1$ is a machine-dependent constant.
The first term in the brackets corresponds to the time spent to solve the heat conductivity problem,
the second and the third terms correspond to tracing of incoming and outcoming light rays respectively.
Decreasing any of these terms separately from the other two terms does not give any significant gain in the performance time,
but often leads to significant loss of accuracy of the problem.
Therefore we decide to dedicate a comparable amount of computation time to all three parts of the algorithm, and to take
\begin{equation}
 N_\mathrm{vis}=N_\mathrm{IR}=N_r^3.
 \label{N}
\end{equation}
Then the computation time is
\begin{equation}
 t_\mathrm{comp} = C_2 t_\mathrm{eq}s N_r^5,
 \label{t_comp2}
\end{equation}
with $C_2$ being another constant.

The error of $p_x$ due to the shot noise is inversely proportional to the square root of the total number of rays emitted,
\begin{equation}
 \Delta p_x \propto \frac{1}{\sqrt{sN_r^2 N_\mathrm{IR}}}.
 \label{delta_p_x}
\end{equation}
Equation \ref{N} provides $\Delta p_x \propto t_\mathrm{comp}$.

In addition to the random error $\Delta p_x$, the simulation has some systematic error due to limited resolution,
and several conditions must be fulfilled for stability and good convergence of the simulation.
We use these conditions to determine the best parameters for each simulation,
and there is always a trade-off between all these conditions.

First of all, the spatial discretization used for our simulation must be relatively fine,
\begin{equation}
 N_r \gg 1
 \label{condition_N_r}.
\end{equation}
Otherwise the model is too rough and gives a bad approximation of the real temperature distribution and thus the real TYORP force.

Secondly, our method of solving the heat conductivity equation requires
\begin{equation}
 s \theta^2 r^2 \gg 1.
 \label{condition_step}
\end{equation}
Our explicit time integration scheme is only conditionally stable,
and when the left-hand side of Equation \ref{condition_step} is of order of unity it loses stability.
Even when the numeric scheme is stable, smaller values of this left-hand side lead to loss of accuracy,
that is why it is warranted to have this as big as possible.

Thirdly, the time allowed for equilibrization of the temperature distribution inside the stone must be large enough 
to allow for the heat wave to cross the stone,
\begin{equation}
 \frac{t_\mathrm{eq}}{\theta r} \gg 1.
 \label{condition_t}
\end{equation}
If this condition is not fulfilled, the temperature distribution inside the stone does not have enough time to reach its periodic diurnal cycle,
and the obtained TYORP force can differ significantly from reality.

It is also necessary that the total energy absorbed or emitted in one time step
is smaller than the heat energy of the outer shell of volume elements.
\begin{equation}
 s N_r \theta^2 r \gg 1.
 \label{condition_T}
\end{equation}
If this condition is not met, then the temperatures of the outmost volume elements oscillate largely during each timestep,
deteriorating the accuracy of the obtained solution.
The total number of rays is $N_\mathrm{vis} \approx N_\mathrm{IR} \approx N_r^3$, 
implying that each volume element at the surface of the stone absorbs and emits of order of $N_r$ rays at each timestep.
As $N_r \gg 1$ from Equation \ref{condition_N_r} we do not expect any great deviations from the mean energy absorbed or emitted, 
and if Equation \ref{condition_T} provides moderate temperature changes for all volume elements on average, 
it also provides moderate temperature differences for each volume element.
But in the whole area of interest condition Equation \ref{condition_T} is weaker than Equation \ref{condition_step} and thus can be neglected.

The three inequalities in Equations \ref{condition_N_r}, \ref{condition_step}, and \ref{condition_t} must hold simultaneously.
It is impossible to select one set of free parameters that provides these conditions in the whole area where we simulate the effect,
and even if this could be done it would be overkill.
So we select the simulation parameters separately for each set of physical parameters. 
If the computation time given by Equation \ref{t_comp2} is constant,
then an increase in the left-hand side of any of the three inequalities can be attained only at the cost of a decrease in the left-hand sides of the others.
We consider these inequalities to have roughly equal importance for the accuracy of the final result,
and thus require the left-hand sides of Equations \ref{condition_N_r}, \ref{condition_step}, and \ref{condition_t} to be equal,
\begin{equation}
 N_r = s \theta^2 r^2 = \frac{t_\mathrm{eq}}{\theta r}.
 \label{equal_separation_of_resources}
\end{equation}
We fix the computation time, and then Equations \ref{N}, \ref{equal_separation_of_resources}, and \ref{t_comp2}
provide us with a system of 5 equations, 
from which we express the 5 simulation parameters $N_r$, $N_\mathrm{vis}$, $N_\mathrm{IR}$, $s$, and $t_\mathrm{eq}$.
Thus for each point of the plot our program automatically determines the simulation parameters from the given simulation time.
We prescribe the same computation time $t_\mathrm{comp}$ to all points in the plot,
so that the shot noise for all the points is limited to nearly the same amount.

The left-hand sides of Equations \ref{condition_N_r}, \ref{condition_step}, \ref{condition_t}, and \ref{condition_T} 
are shown in Figure \ref{error}.
We see that even though we try to choose the values in the first 3 panels 
(left-hand sides of Equations \ref{condition_N_r}, \ref{condition_step}, \ref{condition_t}) equal, they are actually not.
The most important reason is that $t_\mathrm{eq}$ must be integer, and thus not less than 1.
This limitations causes us to allow a bigger value in the 2nd panel, and thus lower values in the 1st and 3rd panels.
Almost everywhere in the plots all the parameters are bigger than 10, 
and go down to about 3 only in the lower left corner of the plot,
where both TYORP and Yarkovsky are negligible anyway.
This figure validates the applicability of our program to the simulations performed.

\section{TYORP of a 3-axial ellipsoid}

In this appendix we compute TYORP dimensionless torque for sinusoidal dependence of $p_x$ on latitude and then estimate the torque for $p_x$ independent of latitude.
The two results are relatively close to each other, and the correct TYORP should lie somewhere between them.

Let us first assume a sinusoidal latitude dependence, 
so that the TYORP stress at each point of the surface of an asteroid is expressed by the formula
\begin{equation}
P=\frac{\Phi}{c} p_0 f \cos \psi,
\label{app_sigma}
\end{equation}
where $p_0$ is $p_x$ at the equator and $f$ is the fraction of the surface occupied by stones.
Let us compute the TYORP torque experienced by the asteroid.

First we consider a surface element $\mathbf{dS}=(dS_x,dS_y,dS_z)$ that has the radius-vector $\mathbf{r}=(r_x,r_y,r_z)$.
The TYORP force acting on the element is tangential to the surface and perpendicular to the rotation axis $\mathbf{e}_z$ of the asteroid,
therefore it is parallel to $\mathbf{e}_z\times\mathbf{dS}=(-dS_y,dS_x,0)$. 
(We assume the following sign convention: $P$ is positive if the force accelerates the asteroid's rotation,
and $\mathbf{e}_z$ is co-directional with the angular velocity of the asteroid.)
Therefore the TYORP force acting on the surface element is
\begin{equation}
 \mathbf{dF}=P\,dS\frac{(-dS_y,dS_x,0)}{\sqrt{dS_x^2+dS_y^2}}.
\end{equation}
The TYORP torque acting on the asteroid is
\begin{eqnarray}
 T_z&=&\oint\,\mathbf{e}_z\cdot[\mathbf{r}\times\mathbf{dF}]= \nonumber \\
 &=&\oint\,(0,0,1)\cdot\left[(r_x,r_y,r_z)\times P\,dS\frac{(-dS_y,dS_x,0)}{\sqrt{dS_x^2+dS_y^2}}\right]= \nonumber \\
 &=&\frac{\Phi}{c} p_0 f \oint\,\cos \psi\,\frac{dS}{\sqrt{dS_x^2+dS_y^2}}(r_x dS_x+r_y dS_y)= \nonumber \\
 &=&\frac{\Phi}{c} p_0 f \oint(r_x dS_x+r_y dS_y).
\end{eqnarray}
The corresponding non-dimensional torque is
\begin{equation}
 \tau_z=\frac{cT_z}{\Phi r_\mathrm\mathrm{eq}^3}=\frac{p_0 f}{r_\mathrm\mathrm{eq}^3}\oint(r_x dS_x+r_y dS_y),
 \label{app_tau}
\end{equation}
where $r_\mathrm\mathrm{eq}$ is volume-equivalent radius of the asteroid.

An important consequence of this formula is that whenever we stretch the asteroid in either polar or equatorial direction, 
its dimensionless TYORP stays unchanged.
Indeed, let say we apply the transformation $(r_x,r_y,r_z)\rightarrow(ar_x,br_y,cr_z)$.
Then $(dS_x,dS_y,dS_z)\rightarrow(bc\,dS_x,ac\,dS_y,ab\,dS_z)$, the integral gets multiplied by the factor $abc$, 
but $r_\mathrm\mathrm{eq}^3$ gets multiplied by the same factor, and the two factors cancel.

In particular, $\tau_z$ for a triaxial ellipsoid rotating around one of its major axes is the same as for a sphere.
The latter is
\begin{eqnarray}
 \tau_z&=&p_0 f\int_0^{2\pi}\,d\phi\int_{-\frac{\pi}{2}}^{\frac{\pi}{2}}\,d\theta(\cos \theta\,\cos \phi\cdot\cos ^2\theta\,\cos \phi+ \nonumber \\
 &+&\cos \theta\,\sin \phi\cdot\cos ^2\theta\,\sin \phi)= \nonumber \\
 &=&2\pi p_0 f\int_{-\frac{\pi}{2}}^{\frac{\pi}{2}}\,d\theta\,\cos ^3\theta=\frac{8\pi}{3}p_0 f \approx 8.38 p_0 f.
 \label{app_sphere}
\end{eqnarray}

\begin{figure*}
  \centering    
  \includegraphics{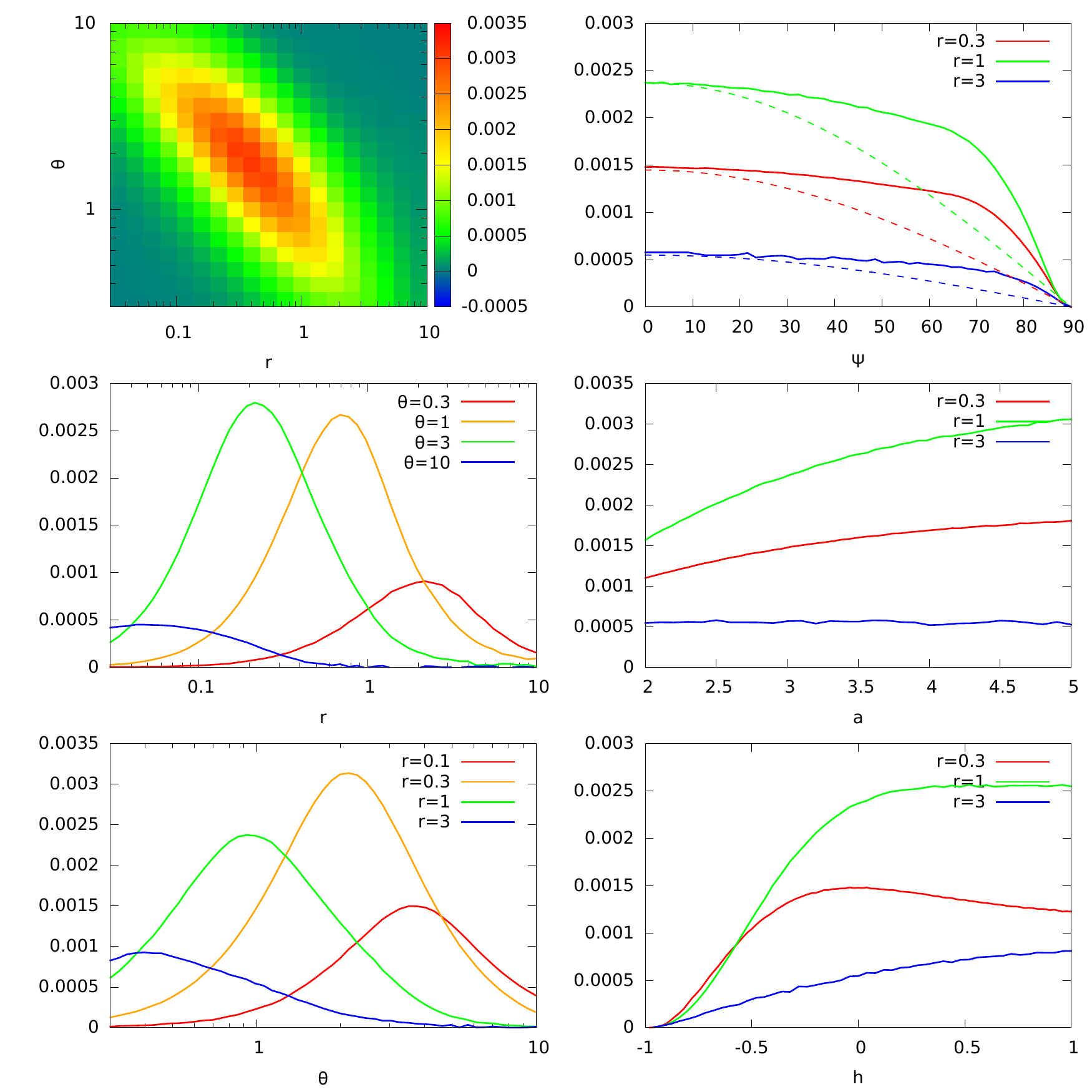}
  \caption{TYORP drag $p_x$ as a function of relevant parameters (see Table \ref{tab-var}). 
  All panels include the point $r=1$, $a=3$, $h=0$, $\psi=0$, $\theta=1$,
  and also one parameter varies along the $x$-axis and another parameter varies between the plotted lines.
  The upper left panel is a colour map showing $p_x$ as a function of $r$ and $\theta$.
  Dashed lines in the upper right panel are cosine functions shown for comparison.}
  \label{p_x}
\end{figure*}

\begin{figure*}
  \centering    
  \includegraphics{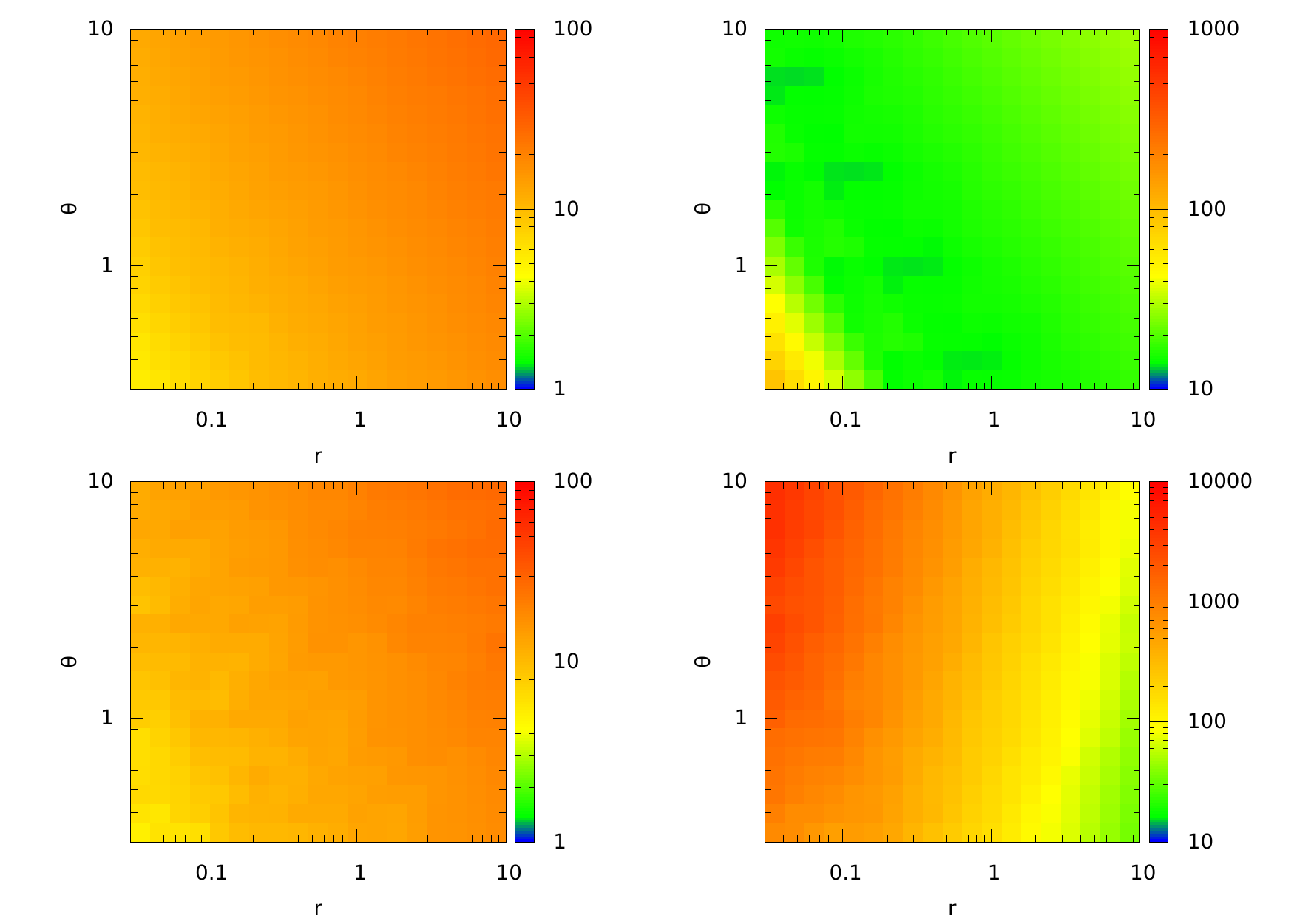}
  \caption{Accuracy test of our code.
  The four panels show left-hand sides of Equations \ref{condition_N_r}, \ref{condition_step}, \ref{condition_t}, and \ref{condition_T}.
  For the good accuracy of the program they all must be much bigger than 1 in the whole area of simulations.}
  \label{error}
\end{figure*}

From Figure \ref{p_x} we see that Equation \ref{app_sigma} is only a rough approximation.
Another extreme could be to say that $P$ does not depend on the latitude at all,
\begin{equation}
P=\frac{\Phi}{c} p_0 f,
\label{app_sigma_2}
\end{equation}
In this approximation the integration for a 3-axial ellipsoid can not be done that easily, 
but we also do not expect dimensionless TYORP $\tau_z$ to significantly depend on the shape.
However, calculation of the torque for a sphere is easy, and it gives
\begin{eqnarray}
 \tau_z=\pi^2 p_0 f \approx 9.87 p_0 f.
 \label{app_sphere_2}
\end{eqnarray}

The coefficients in Equations \ref{app_sphere} and \ref{app_sphere_2} are close to each other,
that gives us a good reason to believe that for an asteroid of any ellipsoidal shape $\tau_z \approx 9 p_0 f$.

The closeness of the results obtained for such different latitude dependencies of $P$ is not surprizing,
as high latitudes give only a minor contribution to the total torque,
firstly because of their small surface area,
and secondly because their small lever arm.
$\tau_z$ is predominantly detemined by low latitudes,
where Equations \ref{app_sigma_2} and \ref{app_sigma} are close to each other.

\begin{table}
\caption{Used notations.} 
\begin{tabular}{c|l} \hline
Notation & Meaning
         \\ \hline
\multicolumn{2}{c}{\textit{geometric configuration}} \\ [-0.1cm]
$R$ & radius of the stones \\
$r$ & dimensiomless radius of the stones $R/L_\mathrm{cond}$\\
$a$ & relative distance between the stones expressed in terms of the stone's radius\\
$h$ & relative height of the center of a stone above the ground in terms of the stone's radius\\
$\psi$ & latitude on the surface \\ 
\multicolumn{2}{c}{\textit{thermal properties}} \\ [-0.1cm]
$C$ & heat capacity of the stone \\
$\rho$ & density of the stone \\
$\kappa$ & heat conductivity of the stone \\
$\sigma$ & Stefan--Boltzmann’s constant \\
$\epsilon$ & heat emissivity \\
$\Phi$ & solar energy flux \\
$A$ & albedo of the stone \\
$\theta$ & thermal parameter \\
\multicolumn{2}{c}{\textit{variables}} \\ [-0.1cm]
$x_i$ & coordinates ($i$=1,2,3) \\
$\xi_i$ & normalized coordinates ($i$=1,2,3) \\
$t$ & time \\
$\phi$ & rotation phase $\omega t$ \\
$T$ & temperature \\
$\tau$ & normalized temperature \\
\multicolumn{2}{c}{\textit{simulation parameters}} \\ [-0.1cm]
$N_r$ & number of nodes along the radius \\
$N_\mathrm{vis}$ & number of incoming rays per step \\
$N_\mathrm{IR}$ & number of outcoming rays per step \\
$s$ & parameter determining the duration of the step \\
$t_\mathrm{eq}$ & number of rotation periods for equilibrization \\
\hline
\end{tabular}
\label{tab-var}
\end{table}

\begin{table}
\caption{Predicted and observed YORP acceleration of Itokawa.
The normalized YORP torque $\tau_z$ is calculated with Equation \ref{tau_z} 
assuming Itokawa's principal axes to be 535, 294, and 209 metres, the mass to be $3.51\times 10^{10}$ kilograms \citet{fujiwara06},
the solar constant $\Phi=1360$ W m$^{-2}$, and the semimajor axis 1.324 AU.}
\begin{tabular}{l|l|l} \hline
Source & $d\omega /dt$ & $\tau_z$
         \\ \hline
Theory: & &\\
\cite{scheeres07} & $-(2.5 \div 4.5)\times 10^{-17}$ rad s$^{-2}$ & $-(0.0015\div 0.0028)$ \\
\cite{durech08ito} & $-(0.730 \div 3.097)\times 10^{-7}$ rad day$^{-2}$ & $-(0.0006 \div 0.0026)$ \\
\cite{breiter09} & $-(2.5 \div 5.5)\times 10^{-7}$ rad day$^{-2}$ & $-(0.0021 \div 0.0046)$ \\
         \hline
Observations: & &\\
\cite{lowry14} & $(3.54 \pm 0.38)\times 10^{-8}$ rad day$^{-2}$ & $0.00029 \pm 0.00003$ \\
         \hline
TYORP (maximum) & & 0.002 \\
\hline
\end{tabular}
\label{tab-itokawa}
\end{table}

\end{document}